\documentclass[11pt]{article} 
\newcommand{\be}{\begin{equation}}
\newcommand{\ee}{\end{equation}}
\newcommand{\bea}{\begin{eqnarray}}
\newcommand{\eea}{\end{eqnarray}}
\newcommand{\sn}{{\rm sn}}

\newcommand{\dn}{{\rm dn}}
\newcommand{\cn}{{\rm cn}}
\newcommand{\sech}{{\rm sech}}

\begin{document}
\vspace{.5in} 
\begin{center} 
{\LARGE{\bf Domain Wall and Periodic Solutions of Coupled $\phi^4$ Models 
in an External Field}}
\end{center} 

\vspace{.3in}
\begin{center} 
{\LARGE{\bf Avinash Khare}} \\ 
{Institute of Physics, Bhubaneswar, Orissa 751005, India}
\end{center} 

\begin{center} 
{\LARGE{\bf Avadh Saxena}} \\ 
{Theoretical Division and Center for Nonlinear Studies, Los
Alamos National Laboratory, Los Alamos, NM 87545, USA}
\end{center} 

\vspace{.9in}
{\bf {Abstract:}}  

Coupled double well ($\phi^4$) one-dimensional potentials abound in 
both condensed matter physics and field theory.  Here we provide an 
exhaustive set of exact periodic  solutions of a coupled $\phi^4$ model 
in an external field in terms of elliptic functions (domain wall arrays) 
and obtain single domain wall solutions in specific limits.  We also 
calculate the energy and interaction between solitons for various 
solutions.  Both topological and nontopological (e.g. some pulse-like 
solutions in the presence of a conjugate field) domain walls are obtained.  
We relate some of these solutions to the recently observed magnetic 
domain walls in certain multiferroic materials and also in the field 
theory context wherever possible.  Discrete analogs of these coupled 
models, relevant for structural transitions on a lattice, are also 
considered.   

\newpage 
  
\section{Introduction} 
There are many physical situations, both in condensed matter and field 
theory, where two double well potentials model the phenomena of interest 
with a specific coupling between the two fields.  One such phenomenon 
of current intense interest is the coexistence of magnetism and 
ferroelectricity (i.e. magnetoelectricity) in a given material.  This 
is a highly desired functionality in technological applications involving 
cross-field response, switching and actuation.  In general, this 
phenomenon is referred to as multiferroic behavior \cite{multif}.  
Recently, two different classes of (single phase) multiferroics, namely 
the orthorhombically distorted perovskites \cite{kimura} and rare earth 
hexagonal structures \cite{fiebig}, have emerged.  The latter show 
magnetic domain walls in the basal planes which can be modeled by a 
coupled $\phi^4$ model \cite{curnoe} in the presence of a magnetic 
field.  Coupled $\phi^4$ models \cite{aubry,abel,jcp} also arise in the 
context of many ferroelectric and other second order phase transitions.  
The coupled $\phi^4$ model for multiferroics \cite{curnoe} has a 
biquadratic coupling whereas the coupled $\phi^4$ model for a surface 
phase transition with hydration forces \cite{jcp}, relevant in biophysics 
context, has a bilinear coupling.  Other types of couplings are also 
known for structural phase transitions with  strain \cite{das}.  

Similarly, there are analogous coupled models in field theory 
\cite{raja,pla}.  Several related models have been discussed in the 
literature and their soliton solutions have been found 
\cite{lai,rao,wang,huang,bazeia,zhu,lou,cao} including periodic ones 
\cite{li,liu,llw}.  Here our motivation is to obtain various possible 
domain wall solutions of these models with either a bilinear or a 
biquadratic coupling and then connect to experimental observations 
wherever possible.  

The paper is organized as follows. In Sec. II we provide the solutions 
for the coupled $\phi^4$ model with an explicit biquadratic coupling in 
the presence of an external field (with an additional linear-quadratic 
coupling) and calculate their energy as well as interaction between the 
solitons.  We also obtain solutions in the limit of no field.  In Sec. 
III we consider a coupled $\phi^4$ model with a bilinear coupling.  In 
Sec. IV we obtain several solutions of a coupled discrete $\phi^4$ model 
with biquadratic coupling (but without an external field). In Sec. V we
obtain a solution of the coupled discrete $\phi^4$ model with bilinear
coupling. To the best of our knowledge, no solution of the coupled 
$\phi^4$ model in an external field is known. Similarly, no solutions 
of either the coupled continuum $\phi^4$ model with a bilinear coupling 
or the discrete case are known.  Even for the continuum coupled case with 
a biquadratic coupling and zero external field, out of the six possible 
solutions only three were known previously \cite{liu} but the other 
three are new.  Finally, we conclude in Sec. VI with remarks on related 
models. 

\section{Coupled $\phi^4$ solutions in an external field}

We consider several exact solutions of the coupled $\phi^4$ system in 
a magnetic field ($H_z$) as given in Ref. \cite{curnoe} for hexagonal  
multiferroics. In particular, there are nine periodic solutions (valid 
for arbitrary $m$, the modulus of elliptic functions), which at $m=1$ 
reduce to just four solutions. In particular, there is one ``bright-bright", 
one ``dark-dark" and one each of bright-dark and dark-bright solutions. 
Notice that the equations of motion are asymmetric in the two scalar 
fields ($\phi$ and $\psi$) due to different coupling of the scalar 
fields to the magnetic field.  Thus, the dark-bright and the bright-dark 
solutions are distinct. 

The potential, with a {\it biquadratic} coupling between the two fields 
and in an external magnetic field ($H_z$), is given by  
\be\label{1}
V=\alpha_1 \phi^2 +\beta_1 \phi^4 +\alpha_2 \psi^2 +\beta_2 \psi^4 
+\gamma\phi^2 \psi^2
-H_z[\rho_1 \phi+\rho_2 \phi^3+\rho_3 \phi \psi^2]\,, 
\ee
where $\alpha_i$, $\beta_i$, $\gamma$ and $\rho_i$ are material (or 
system) dependent parameters.  Hence the (static) equations of motion are  
\be\label{2}
\frac{d^2 \phi}{dx^2}=2\alpha_1 \phi +4\beta_1 \phi^3 
+2\gamma\phi \psi^2
-H_z[\rho_1+3\rho_2 \phi^2+\rho_3 \psi^2]\,,
\ee
\be\label{3}
\frac{d^2 \psi}{dx^2}=2\alpha_2 \psi +4\beta_2 \psi^3 
+2\gamma\phi^{2} \psi
-2H_z\rho_3 \phi \psi\,.
\ee
These coupled set of equations admit several periodic solutions which 
we now discuss one by one systematically.

For static solutions the energy is given by
\be
E=\int \left[\frac{1}{2}\left(\frac{d\phi}{dx}\right)^2+\frac{1}{2} 
\left(\frac{d\psi}{dx}\right)^2 +V(\phi,\psi)\right]\,dx \,,
\ee
where the limits of integration are from $-\infty$ to $\infty$ in the
case of hyperbolic solutions (i.e., single solitons) on the full line. 
On the other hand, in the case of periodic solutions (i.e. soliton 
lattices), the limits are from $-K(m)$ to $+K(m)$.  Here $K(m)$ [and 
$E(m)$ below] denote the complete elliptic integral of the first (and 
second) kind \cite{GR}. Using equations of motion, one can show that 
for all of our solutions
\be
V(\phi,\psi)=\left[\frac{1}{2}\left(\frac{d\phi}{dx}\right)^2
+\frac{1}{2}\left(\frac{d\psi}{dx}\right)^2\right] +C \,,
\ee
where the constant $C$ in general varies from solution to solution.
Hence the energy $\hat{E}=E-\int C\,dx$ is given by
\be
\hat{E} \equiv E-\int C\,dx =\int\left[\left(\frac{d\phi}{dx}\right)^2 
+\left(\frac{d\psi}{dx}\right)^2\right]\,dx .  
\ee
Below we will give explicit expressions for energy in the case of all 
nine periodic solutions (and hence the corresponding four hyperbolic 
solutions). In each case we also provide an expression for the constant $C$.

\subsection{Solution I}

We look for the most general periodic solutions in terms of the Jacobi 
elliptic functions sn($x,m$), cn($x,m$) and dn($x,m)$ \cite{GR}.  It is 
not difficult to show that
\be\label{1r}
\phi =F+A\sn[D(x+x_0),m]\,,~~\psi =G+B\sn[D(x+x_0),m]\,,
\ee
is an exact solution provided the following eight coupled equations are 
satisfied
\be
2\alpha_1 F+4\beta_1 F^3+2\gamma FG^2-H_z\rho_1-3H_z \rho_2F^2
-H_z\rho_3G^2 =0\,,
\ee
\be
2\alpha_1 A+12\beta_1 F^2 A+4\gamma BFG+2\gamma AG^2
-6H_z \rho_2 AF-2H_z\rho_3 BG =-(1+m)AD^2\,,
\ee
\be
12\beta_1 F A^2+2\gamma FB^2+4\gamma ABG-3H_z \rho_2A^2
-H_z\rho_3B^2 =0\,,
\ee
\be
2\beta_1 A^2+\gamma B^2=mD^2\,,
\ee
\be 
2\alpha_2 G+4\beta_2 G^3+2\gamma GF^2-2H_z\rho_3GF =0\,,
\ee
\be
2\alpha_2 B+12\beta_2 G^2 B+4\gamma AFG+2\gamma BF^2
-2H_z\rho_3 (BF+AG) =-(1+m)BD^2\,,
\ee
\be
12\beta_2 G B^2+2\gamma GA^2+4\gamma ABF
-2H_z\rho_3AB =0\,,
\ee
\be
2\beta_2 B^2+\gamma A^2=mD^2\,.
\ee
Here $A$ and $B$ denote the amplitudes of the ``kink lattice", $F$ 
and $G$ are constants, $D$ is an inverse characteristic length and 
$x_0$ is the (arbitrary) location of the kink.  Five of these equations 
determine the five unknowns $A,B,D,F,G$ while the other three equations, 
give three constraints between the nine parameters $\alpha_{1,2},\beta_{1,2}, 
\gamma,H_z,\rho_1, \rho_2,\rho_3$. 
In particular, $A$ and $B$ are given by
\be\label{1.1}
A^2=\frac{mD^2[2\beta_2-\gamma]}{[4\beta_1 \beta_2 -\gamma^2]}\,, ~~~ 
B^2=\frac{mD^2[2\beta_1-\gamma]}{[4\beta_1 \beta_2 -\gamma^2]}\,. 
\ee
Thus this solution exists provided $2\beta_1>\gamma$ and $2\beta_2>\gamma$. 

It may be noted here that in case both $F,G=0$ then no solution exists
so long as $H_z \ne 0$. In fact this is true for all the nine solutions 
that we discuss below. There are two special cases (i.e., when either 
$F=0$ or $G=0$) when the analysis becomes somewhat simpler.

\noindent{\bf G=0, F$\ne$0:}

In this case $A,B$ are again given by Eq. (\ref{1.1}) while
\be\label{1.2}
(1+m)D^2=\frac{\rho_2 \rho_3 H_z^2}{2\gamma}-\frac{2\rho_1
\gamma}{\rho_3}\,,~~F=\frac{\rho_3H_z}{2\gamma}\,.
\ee
The three (and not four, since one of the equations is identically 
satisfied) constraints are
\be\label{1.3}
\beta_1=\frac{\rho_2 \gamma}{2 \rho_3}\,,~~
\alpha_1=\frac{\rho_2 \rho_3 H_z^2}{2\gamma}+\frac{\rho_1
\gamma}{\rho_3}\,,~~
\alpha_2=\frac{\rho_3^2 H_z^2}{4\gamma}-
\frac{\rho_2 \rho_3 H_z^2}{4\gamma}+\frac{\rho_1
\gamma}{\rho_3}\,.
\ee

\noindent{\bf m=1:}

In this limiting case we have a bright-bright soliton solution given by
\be\label{1.4}
\phi =F+A\tanh[D(x+x_0)]\,,~~\psi =B\tanh[D(x+x_0)]\,,
\ee

with $A$, $B$ and $D$ determined by  
\be\label{1.5}
A^2=\frac{D^2[2\beta_2-\gamma]}{[4\beta_1 \beta_2 -\gamma^2]}\,,
~B^2=\frac{D^2[2\beta_1-\gamma]}{[4\beta_1 \beta_2 -\gamma^2]}\,,
~2D^2=\frac{\rho_2 \rho_3 H_z^2}{2\gamma}-\frac{2\rho_1
\gamma}{\rho_3}\,,
\ee
while the other relations remain unchanged and are again given by 
Eqs.  (\ref{1.2}) and (\ref{1.3}).

\noindent{\bf F=0, G$\ne$0:}

In this case $A,B$ are again given by Eq. (\ref{1.1}) with
\be
(1+m)D^2=4\alpha_2+\frac{2 \rho_3 H_z AG}{B}\,,~~
G^2=-\frac{\alpha_2}{2\beta_2}\,,
\ee
and the corresponding four constraints are
\bea
&&\beta_2=\frac{\rho_3 \alpha_2}{2 \rho_1}\,,~~
(1+m)D^2=-2\alpha_1-2\gamma G^2+\frac{2 \rho_3 H_z BG}{A}\,,
\nonumber \\
&&4\gamma ABG=3H_z\rho_2 A^2+H_z\rho_3 B^2\,,~~
\rho_3 H_zAB=\gamma G A^2+6\beta_2 GB^2\,.
\eea
The solution at $m=1$ can be easily written down as above.

\noindent{\bf Special case of $H_z=0$:}

In this case the field equations are completely symmetric in the two
fields $\phi,\psi$ and solution exists even if both $F,G=0$. In
particular, with $H_z=0$, the solution is as given by Eq. (7) 
but with $F=0=G$ \cite{liu} where $A,B$ are again as given by Eq. 
(16) and further both $\alpha_1,\alpha_2$ turn out to be negative,
i.e.
\be
\alpha_1=\alpha_2=-\frac{(1+m)D^2}{2}\,.
\ee

\noindent{\bf Special case of $\gamma^2 = 4\beta_1 \beta_2$}

One can show that the solution (\ref{1r}) exists even in case $\gamma^2=
4\beta_1 \beta_2$. It turns out  such a solution
exists only if 
\be
2\beta_1=2\beta_2=\gamma\,,
\ee
and that in this case one cannot determine $A,B$. However, they must
satisfy the constraint
\be\label{1c}
A^2+B^2=\frac{mD^2}{\gamma}\,.
\ee
Other relations can be easily worked out depending on if $F$ or $G$ 
(or neither) is zero. For example, in case $G=0,F \ne 0$, one has
\be
\rho_2 = \rho_3\,,~~\alpha_2=\frac{\rho_1 \gamma}{\rho_3}\,,~~
\alpha_1-\alpha_2 =\frac{\rho_3^2 H_z^2}{2\gamma}\,.
\ee

\noindent{\bf Energy:} Corresponding to the periodic solution [Eq. 
(\ref{1r}) with $G=0$] 
the energy $\hat{E}$ and the constant $C$ are given by 
\bea
&&\hat{E}=\frac{2(A^2+B^2)D}{3m}[(1+m)E(m)-(1-m)K(m)]\,, \nonumber \\
&&C=-\frac{1}{2}(A^2+B^2)D^2-F^2[\alpha_1+3\beta_1 F^2-2H_z \rho_2 F]\,.
\eea

It is worth pointing out that even in the case of the solution
(\ref{1r}) with either $F=0$, $G \ne 0$ or both $F,G$ nonzero, 
the energy $\hat{E}$ is the same. Only the value of $C$ is different. For
example, in the case of $F=0$, $G \ne 0$, $C$ is given by 
\be
C=-\frac{1}{2}(A^2+B^2)D^2-\frac{\alpha_2^2}{4\beta_2}\,, 
\ee
while in the case of both $F$, $G$ being nonzero, $C$ is 
\be
C=-\frac{1}{2}(A^2+B^2)D^2-F^2[\alpha_1+3\beta_1 F^2-2H_z \rho_2 F]\,
-G^2[\alpha_2+3\beta_2 G^2+3\gamma F^2-2H_z \rho_3 F]\,.
\ee

On using the expansion formulas for $E(m)$ and $K(m)$ around $m=1$ as
given in \cite{GR}
\be
K(m)=\ln\left(\frac{4}{\sqrt{1-m}}\right)+\frac{(1-m)}{4}\left[\ln\left( 
\frac{4}{\sqrt{1-m}}\right)-1\right]+...
\ee
\be
E(m)=1+\frac{(1-m)}{2}\left[\ln\left(\frac{4}{\sqrt{1-m}}\right)-\frac{1}{2} 
\right]+..., 
\ee
for $m$ near one, the energy of the periodic solution can be rewritten 
as the energy of the corresponding hyperbolic (bright-bright) soliton 
solution [Eq. (\ref{1.4})]  
plus the interaction energy. We find
\be
\hat{E}=E_{kink}+E_{int}=(A^2+B^2)D \left[\frac{4}{3}+\frac{(1-m)}{3}\right]\,.
\ee
Note that this solution exists only when $2\beta_1 \ge \gamma$, $2\beta_2
\ge \gamma$ and $4\beta_1 \beta_2 \ge \gamma^2$.  The interaction energy 
vanishes at exactly $m=1$, as it should.  

\subsection{Solution II}

A different type of solution (``pulse lattice") is given by 
\be\label{2.1}
\phi =F+A\cn[D(x+x_0),m]\,,~~\psi =G+B\cn[D(x+x_0),m]\,,
\ee
provided the following eight coupled equations are satisfied
\be
2\alpha_1 F+4\beta_1 F^3+2\gamma FG^2-H_z\rho_1-3H_z \rho_2F^2
-H_z\rho_3G^2 =0\,,
\ee
\be
2\alpha_1 A+12\beta_1 F^2 A+4\gamma BFG+2\gamma AG^2
-6H_z \rho_2 AF-2H_z\rho_3 BG =(2m-1)AD^2\,,
\ee
\be
12\beta_1 F A^2+2\gamma FB^2+4\gamma ABG-3H_z \rho_2A^2
-H_z\rho_3B^2 =0\,,
\ee
\be
2\beta_1 A^2+\gamma B^2=-mD^2\,,
\ee
\be
2\alpha_2 G+4\beta_2 G^3+2\gamma GF^2-2H_z\rho_3GF =0\,,
\ee
\be
2\alpha_2 B+12\beta_2 G^2 B+4\gamma AFG+2\gamma BF^2
-2H_z\rho_3 (BF+AG) =(2m-1)BD^2\,,
\ee
\be
12\beta_2 G B^2+2\gamma GA^2+4\gamma ABF
-2H_z\rho_3AB =0\,,
\ee
\be
2\beta_2 B^2+\gamma A^2=-mD^2\,.
\ee
Notice that two of these equations are meaningful only if $\gamma<0$ 
since $\beta_1,\beta_2 >0$ from stability considerations.  Thus, we 
write $\gamma=-|\gamma|$.  Five of these equations determine the five  
unknowns $A,B,D,F,G$ while other three equations, give three constraints 
between the nine parameters $\alpha_{1,2},\beta_{1,2}, \gamma,H_z,\rho_1,
\rho_2,\rho_3$.
In particular, $A$ and $B$ are given by
\be\label{2.2}
A^2=\frac{mD^2[2\beta_2+|\gamma|]}{[\gamma^2-4\beta_1 \beta_2]}\,, ~~~ 
B^2=\frac{mD^2[2\beta_1+|\gamma|]}{[\gamma^2-4\beta_1 \beta_2]}\,.  
\ee
Thus this solution exists provided $\gamma^2 > 4\beta_1 \beta_2$. 

There are two special cases when the analysis becomes somewhat simpler
and we consider both the cases one by one.

\noindent{\bf G=0, F$\ne$0:}

In this case $A,B$ are again given by Eq. (\ref{2.2})  
while 
\be\label{2.3}
(2m-1)D^2=\frac{\rho_2 \rho_3 H_z^2}{2|\gamma|}-\frac{2\rho_1
|\gamma|}{\rho_3}\,,~~F=-\frac{\rho_3H_z}{2|\gamma|}\,,
\ee
and the corresponding three constraints are
\be\label{2.4}
\beta_1=-\frac{\rho_2 |\gamma|}{2 \rho_3}\,,~~
\alpha_1=-\frac{\rho_2 \rho_3 H_z^2}{2|\gamma|}-\frac{\rho_1
|\gamma|}{\rho_3}\,,~~
\alpha_2=-\frac{\rho_3^2 H_z^2}{4|\gamma|}+
\frac{\rho_2 \rho_3 H_z^2}{4|\gamma|}-\frac{\rho_1
|\gamma|}{\rho_3}\,.
\ee

\noindent{\bf F=0, G$\ne$0:}

In this case $A,B$ are again given by Eq. (\ref{2.2}) with
\be\label{2.4x}
(2m-1)D^2=2\alpha_1-\frac{2 \rho_3 H_z BG}{A}-2|\gamma|G^2\,,~~
G^2=-\frac{\alpha_2}{2\beta_2}\,,
\ee
and the corresponding four constraints are
\bea\label{2.4y}
&&\beta_2=\frac{\rho_3 \alpha_2}{2 \rho_1}\,,~~
(2m-1)D^2=-4\alpha_2-\frac{2 \rho_3 H_z AG}{B}\,,
\nonumber \\
&&-4|\gamma| ABG=3H_z\rho_2 A^2+H_z\rho_3 B^2\,,~~
\rho_3 H_zAB=-|\gamma| G A^2+6\beta_2 GB^2\,.
\eea

\noindent{\bf Special case of $H_z$=0:}

In this case the field equations are completely symmetric in the two
fields $\phi,\psi$ and a solution exists even when both $F,G=0$. In
particular, with $H_z=0$, the solution is as given by Eq. (\ref{2.1}) 
but with $F=0=G$ \cite{liu} where $A,B$ are again as given by Eq. 
(\ref{2.2}) and furthermore, $\alpha_1,\alpha_2$ are positive (negative)
so long as $m > (<)$ 1/2, i.e.
\be
\alpha_1=\alpha_2=\frac{(2m-1)D^2}{2}\,.
\ee

\noindent{\bf Energy:} Corresponding to the ``pulse lattice'' solution 
[Eq. (\ref{2.1}) with $G=0$] 
the energy is given by
\bea
&&\hat{E}=\frac{2(A^2+B^2)D}{3m}[(2m-1)E(m)+(1-m)K(m)]\,, \nonumber \\
&&C=-\frac{1}{2}(1-m)(A^2+B^2)D^2-F^2[\alpha_1+3\beta_1 F^2-2H_z \rho_2 F]\,.
\eea

It is worth pointing out that even in the case of the solution
(\ref{2.1}) with either $F=0$, $G \ne 0$ or both $F$, $G$ nonzero, 
the energy $\hat{E}$ is the same. Only the value of $C$ is different. For
example, in the case of $F=0$, $G \ne 0$, $C$ is given by 
\be
C=-\frac{1}{2}(1-m)(A^2+B^2)D^2-\frac{\alpha_2^2}{4\beta_2}\,, 
\ee
while in the case of both $F$, $G$ being nonzero, $C$ is 
\be
C=-\frac{1}{2}(1-m)(A^2+B^2)D^2-F^2[\alpha_1+3\beta_1 F^2-2H_z \rho_2 F]\,
-G^2[\alpha_2+3\beta_2 G^2+3\gamma F^2-2H_z \rho_3 F]\,.
\ee

For $m$ near one, the energy of this periodic solution can
be rewritten as the energy of the corresponding hyperbolic
(dark-dark) soliton solution
\be\label{2t}
\phi =F+A\sech[D(x+x_0)]\,,~~\psi =B\sech[D(x+x_0)]\,,
\ee
plus the interaction energy. We find
\be
\hat{E}=E_{pulse}+E_{int}=(A^2+B^2)D
\left[\frac{2}{3}-\frac{5(1-m)}{6}+(1-m)\ln\left(\frac{4}{\sqrt{1-m}}\right) 
\right]\,.
\ee
Note that this solution exists only when $\gamma < 0, 4\beta_1 \beta_2 < 
\gamma^2$.  Again, the interaction energy vanishes at $m=1$.

\subsection{Solution III}

In this case, there is another ``pulse lattice'' solution which is 
given by 
\be\label{3.1}
\phi =F+A\dn[D(x+x_0),m]\,,~~\psi =G+B\dn[D(x+x_0),m]\,,
\ee
provided the following eight coupled equations are satisfied
\be
2\alpha_1 F+4\beta_1 F^3+2\gamma FG^2-H_z\rho_1-3H_z \rho_2F^2
-H_z\rho_3G^2 =0\,,
\ee
\be
2\alpha_1 A+12\beta_1 F^2 A+4\gamma BFG+2\gamma AG^2
-6H_z \rho_2 AF-2H_z\rho_3 BG =(2-m)AD^2\,,
\ee
\be
12\beta_1 F A^2+2\gamma FB^2+4\gamma ABG-3H_z \rho_2A^2
-H_z\rho_3B^2 =0\,,
\ee
\be
2\beta_1 A^2+\gamma B^2=-D^2\,,
\ee
\be
2\alpha_2 G+4\beta_2 G^3+2\gamma GF^2-2H_z\rho_3GF =0\,,
\ee
\be
2\alpha_2 B+12\beta_2 G^2 B+4\gamma AFG+2\gamma BF^2
-2H_z\rho_3 (BF+AG) =(2-m)BD^2\,,
\ee
\be
12\beta_2 G B^2+2\gamma GA^2+4\gamma ABF
-2H_z\rho_3AB =0\,,
\ee
\be
2\beta_2 B^2+\gamma A^2=-D^2\,.
\ee
Notice that (as in the $\cn-\cn$ case) two of these equations are 
meaningful only if $\gamma < 0$ since $\beta_1,\beta_2 >0$ from 
stability considerations.  We therefore write $\gamma=-|\gamma|$.  Five 
of these equations determine the five unknowns $A,B,D,F,G$ while the 
other three equations give three constraints between the nine parameters 
$\alpha_{1,2},\beta_{1,2}, |\gamma|,H_z,\rho_1, \rho_2,\rho_3$.
In particular, $A$ and $B$ are given by
\be\label{3.2}
A^2=\frac{D^2[2\beta_2+|\gamma|]}{[\gamma^2-4\beta_1 \beta_2]}\,, ~~~ 
B^2=\frac{D^2[2\beta_1+|\gamma|]}{[|\gamma|^2-4\beta_1 \beta_2]}\,.  
\ee

There are two special cases when the analysis becomes somewhat simpler
and we consider both the cases one by one.

\noindent{\bf G=0, F$\ne$0:}

In this case $A,B$ are again given by Eq. (\ref{3.2})  
while 
\be\label{3.3}
(2-m)D^2=\frac{\rho_2 \rho_3 H_z^2}{2|\gamma|}-\frac{2\rho_1
|\gamma|}{\rho_3}\,,~~F=-\frac{\rho_3H_z}{2|\gamma|}\,,
\ee
and the three constraints are
\be\label{3.4}
\beta_1=-\frac{\rho_2 |\gamma|}{2 \rho_3}\,,~~
\alpha_1=-\frac{\rho_2 \rho_3 H_z^2}{2|\gamma|}-\frac{\rho_1
|\gamma|}{\rho_3}\,,~~
\alpha_2=-\frac{\rho_3^2 H_z^2}{4|\gamma|}+
\frac{\rho_2 \rho_3 H_z^2}{4|\gamma|}-\frac{\rho_1
|\gamma|}{\rho_3}\,.
\ee

\noindent{\bf F=0, G$\ne$0:}

In this case $A,B$ are again given by Eq. (\ref{3.2}) with
\be
(2-m)D^2=2\alpha_1-\frac{2 \rho_3 H_z BG}{A}-2|\gamma|G^2\,,~~
G^2=-\frac{\alpha_2}{2\beta_2}\,,
\ee
and the corresponding four constraints are
\bea
&&\beta_2=\frac{\rho_3 \alpha_2}{2 \rho_1}\,,~~
(2-m)D^2=-4\alpha_2-\frac{2 \rho_3 H_z AG}{B}\,,
\nonumber \\
&&-4|\gamma| ABG=3H_z\rho_2 A^2+H_z\rho_3 B^2\,,~~
\rho_3 H_zAB=-|\gamma| G A^2+6\beta_2 GB^2\,.
\eea

\noindent{\bf Special case of $H_z$=0:}

With $H_z=0$, the solution is as given by Eq. (\ref{3.1}) but with 
$F=0=G$ \cite{liu} where $A,B$ are again as given by Eq. (\ref{3.2}) 
and furthermore, both $\alpha_1,\alpha_2$ are positive definite. 
\be
\alpha_1=\alpha_2=\frac{(2-m)D^2}{2}\,.
\ee

\noindent{\bf Energy:} Corresponding to the ``pulse lattice'' solution 
[Eq. (\ref{3.1}) with $G=0$] 
the energy is given by
\bea
&&\hat{E}=\frac{2(A^2+B^2)D}{3}[(2-m)E(m)-(1-m)K(m)]\,, \nonumber \\
&&C=\frac{1}{2}(1-m)(A^2+B^2)D^2-F^2[\alpha_1+3\beta_1 F^2-2H_z \rho_2 F]\,.
\eea

It is worth pointing out that even in the case of the solution
(\ref{3.1}) with either $F=0$, $G \ne 0$ or both $F$, $G$ nonzero, 
the energy $\hat{E}$ is the same. Only the value of $C$ is different. For
example, in the case of $F=0$, $G \ne 0$, $C$ is given by 
\be
C=\frac{1}{2}(1-m)(A^2+B^2)D^2-\frac{\alpha_2^2}{4\beta_2}\,, 
\ee
while in the case of both $F,G$ being nonzero, $C$ is 
\be
C=\frac{1}{2}(1-m)(A^2+B^2)D^2-F^2[\alpha_1+3\beta_1 F^2-2H_z \rho_2 F]\,
-G^2[\alpha_2+3\beta_2 G^2+3\gamma F^2-2H_z \rho_3 F]\,.
\ee

For $m$ near one, the energy of this periodic solution can
be rewritten as the energy of the corresponding hyperbolic
(dark-dark) soliton solution as given by Eq. (\ref{2t})
plus the interaction energy. We find
\be
\hat{E}=E_{pulse}+E_{int}=(A^2+B^2)D
\left[\frac{2}{3}-\frac{(1-m)}{2}-(1-m)\ln\left(\frac{4}{\sqrt{1-m}}\right) 
\right]\,.
\ee
Note that this solution also exists only when $\gamma < 0$, $4\beta_1 
\beta_2 < \gamma^2$.  Again, the interaction energy vanishes at $m=1$.

\subsection{Solution IV}

In addition to the $\cn-\cn$ and $\dn-\dn$ solutions discussed above, 
there are two novel (mixed) soliton solutions of $\dn-\cn$ and $\cn-\dn$ 
type. Let us discuss them one by one.  We shall see that for these two 
solutions (in fact it is true for all the six solutions that we discuss 
below) $G$ is necessarily zero while $F$ is necessarily nonzero (unless
$H_z=0$), otherwise the solution does not exist.  
In particular, it is easily shown that
\be\label{4.1}
\phi =F+A\dn[D(x+x_0),m]\,,~~\psi =G+B\cn[D(x+x_0),m]\,,
\ee
is a solution
provided $G=0$ and further, the following seven coupled equations are 
satisfied
\be
2\alpha_1 F+4\beta_1 F^3+(2/m)\gamma (m-1)FB^2-H_z\rho_1-3H_z \rho_2F^2
+(1/m)(1-m)H_z\rho_3B^2 =0\,,
\ee
\be
2\alpha_1 A+12\beta_1 F^2 A+(2/m)(m-1)\gamma AB^2
-6H_z \rho_2 AF =(2-m)AD^2\,,
\ee
\be
12\beta_1 F A^2+(2/m)\gamma FB^2-3H_z \rho_2A^2
-(1/m)H_z\rho_3B^2 =0\,,
\ee
\be
\gamma B^2+2m\beta_1 A^2=-mD^2\,,
\ee
\be
2\alpha_2 B+2(1-m)\gamma A^2B+2\gamma BF^2
-2H_z\rho_3 BF =(2m-1)BD^2\,,
\ee
\be
2\beta_2 B^2+m \gamma A^2=-mD^2\,,
\ee
\be
4\gamma FAB-2H_z\rho_3AB=0\,.
\ee

Again a solution exists only if $\gamma<0$ and thus we write $\gamma
=-|\gamma|$. 
The solution (\ref{4.1}) exists provided
\be\label{4.2}
A^2=\frac{D^2[|\gamma|+2\beta_2]}{[\gamma^2-4\beta_1 \beta_2]}\,, ~~~ 
B^2=\frac{mD^2[2\beta_1+|\gamma|]}{[\gamma^2-4\beta_1 \beta_2]}\,,  
\ee
where
\be\label{4.3}
(2-m)D^2=\frac{2}{m}(1-m)|\gamma|B^2-\frac{2\rho_1 |\gamma|}{\rho_3}
+\frac{\rho_2 \rho_3 H_z^2}{2|\gamma|}\,,
~~F=-\frac{\rho_3H_z}{2|\gamma|}\,,
\ee
and the three constraints are
\bea
&&\beta_1=-\frac{\rho_2 |\gamma|}{2 \rho_3}\,,~~
\alpha_2=-\frac{\rho_2 \rho_3 H_z^2}{2|\gamma|}-\frac{\rho_1
|\gamma|}{\rho_3}\,, \nonumber \\
&&(2m-1)D^2=2\alpha_2-2(1-m)|\gamma|A^2+\frac{H_z^2
\rho_3^2}{2|\gamma|}\,.
\eea

\noindent{\bf Special case of $H_z=0$:}

With $H_z=0$, the solution is as given by Eq. (\ref{4.1}) but with 
$F=0=G$  where $A$ and $B$ are again as given by Eq. (\ref{4.2}) and 
furthermore, $\alpha_2$ is positive, i.e. 
\be
\alpha_1=\frac{mD^2}{2}-2(1-m)\beta_1 A^2\,,~
m\alpha_2=m\frac{D^2}{2}+2(1-m)\beta_2 B^2 > 0\,.
\ee

\noindent{\bf Energy:} Corresponding to the ``pulse lattice'' solution 
[Eq. (\ref{4.1}) with $G=0$] 
the energy $\hat{E}$ and $C$ are given by
\bea
&&\hat{E}=\frac{2D}{3m}\bigg ([(2-m)mA^2+(2m-1)B^2]E(m)
+(1-m)(B^2-2mA^2)K(m) \bigg )\,, \nonumber \\
&&C=\frac{1}{2}A^2D^2(1-m)-F^2[\alpha_1+3\beta_1 F^2-2H_z \rho_2 F]
+\frac{(1-m)}{m^2}[m\gamma F^2 -\alpha_2 m +\beta_2 (1-m)B^2]B^2\,.
\eea
For $m$ near one, the energy of this periodic solution can
be rewritten as the energy of the corresponding hyperbolic
(dark-dark) soliton solution (\ref{2t})
plus the interaction energy. We find
\be
\hat{E}=E_{pulse}+E_{int}=D\left[\frac{2}{3}(A^2+B^2)
+\frac{(1-m)}{6}(3A^2-5B^2)+(1-m)(B^2-A^2)\ln\left(\frac{4}{\sqrt{1-m}} 
\right) \right]. 
\ee
Note that this solution also exists only when $\gamma < 0, 4\beta_1 
\beta_2 < \gamma^2$.  The interaction energy vanishes for $m=1$, as 
it should.

\subsection{Solution V}

It is easily shown that there is also a $\cn-\dn$ solution which is
distinct from the above $\dn-\cn$ solution. This solution is given by
\be\label{5.1}
\phi =F+A\cn[D(x+x_0),m]\,,~~\psi =G+B\dn[D(x+x_0),m]\,,
\ee
provided $G=0$ and the following seven coupled equations are satisfied
\be
2\alpha_1 F+4\beta_1 F^3+2\gamma (1-m)FB^2-H_z\rho_1-3H_z \rho_2F^2
-(1-m)H_z\rho_3B^2 =0\,,
\ee
\be
2\alpha_1 A+12\beta_1 F^2 A+2(1-m)\gamma AB^2
-6H_z \rho_2 AF =-(1-2m)AD^2\,,
\ee
\be
12\beta_1 F A^2+2m\gamma FB^2-3H_z \rho_2A^2
-mH_z\rho_3B^2 =0\,,
\ee
\be
m\gamma B^2+2\beta_1 A^2=-mD^2\,,
\ee
\be
2\alpha_2 B-(2/m)(1-m)\gamma A^2B+2\gamma BF^2
-2H_z\rho_3 BF =(2-m)BD^2\,,
\ee
\be
2m\beta_2 B^2+\gamma A^2=-mD^2\,,
\ee
\be
4\gamma FAB-2H_z\rho_3AB=0\,.
\ee

Again a solution exists only if $\gamma<0$ and hence we put $\gamma
=-|\gamma|$. The solution turns out to be
\be\label{5.2}
A^2=\frac{mD^2[|\gamma|+2\beta_2]}{[|\gamma|^2-4\beta_1 \beta_2]}\,, ~~~ 
B^2=\frac{D^2[2\beta_1+|\gamma|]}{[|\gamma|^2-4\beta_1 \beta_2]}\,,
\ee
where
\be\label{5.3}
(2m-1)D^2=-2(1-m)|\gamma|B^2-\frac{2\rho_1 |\gamma|}{\rho_3}
+\frac{\rho_2 \rho_3 H_z^2}{2|\gamma|}\,,
~~F=-\frac{\rho_3H_z}{2|\gamma|}\,,
\ee
and the three constraints are
\bea\label{5.4}
&&\beta_1=-\frac{\rho_2 |\gamma|}{2 \rho_3}\,,~~
\alpha_1=-\frac{\rho_2 \rho_3 H_z^2}{2|\gamma|}-\frac{\rho_1
|\gamma|}{\rho_3}\,, \nonumber \\
&&(2-m)D^2=2\alpha_2+(2/m)(1-m)|\gamma|A^2+\frac{H_z^2
\rho_3^2}{2|\gamma|}\,.
\eea

\noindent{\bf Special case of $H_z=0$:}

With $H_z=0$, the solution is as given by Eq. (\ref{5.1}) but with 
$F=0=G$  where $A$ and $B$ are again as given by Eq. (\ref{5.2}) and 
furthermore, $\alpha_1$ is positive, i.e. 
\be
m\alpha_1=m\frac{D^2}{2}+2(1-m)\beta_1 A^2 > 0\,,
\alpha_2=\frac{mD^2}{2}-2(1-m)\beta_2 B^2\,.
\ee
It is worth pointing out that with $H_z=0$, the field equations are
completely symmetric between the two fields $\phi$ and $\psi$ and 
hence the solutions IV and V are identical in the limit $H_z=0$.

\noindent{\bf m=1:}

In the special case of $m=1$ and $G=0$, $F \ne 0$, all the four solutions 
II to V reduce to a dark-dark type soliton solution (\ref{2t}), i.e.
\be\label{5.5}
\phi =F+A\sech[D(x+x_0)]\,,~~\psi =B\sech[D(x+x_0)]\,,
\ee
with $A$, $B$, and $D$ given by
\be\label{5.6}
A^2=\frac{D^2[|\gamma|+2\beta_2]}{[|\gamma|^2-4\beta_1 \beta_2]}\,, ~~~ 
B^2=\frac{D^2[2\beta_1+|\gamma|]}{[|\gamma|^2-4\beta_1 \beta_2]}\,, ~~~ 
D^2=-\frac{2\rho_1 |\gamma|}{\rho_3}
+\frac{\rho_2 \rho_3 H_z^2}{2|\gamma|}\,,
\ee
while the other relations remain unchanged and are again given by 
Eqs.  (\ref{2.3}) and (\ref{2.4}).

Similarly, in the special case of $m=1$ and $F=0$, $G \ne 0$, 
both the solutions II and III 
reduce to a dark-dark type soliton solution
\be\label{5.5a}
\phi =A\sech[D(x+x_0)]\,,~~\psi =G+B\sech[D(x+x_0)]\,,
\ee
with $A$ and $B$ again given by Eq. (\ref{5.6}) while
\be\label{5.6a}
D^2=2\alpha_1-\frac{2 \rho_3 H_z BG}{A}-2|\gamma|G^2\,,~~
D^2 =-4\alpha_2-\frac{2 \rho_3 H_z AG}{B}\,.
\ee
Other relations remain unchanged and are again given by 
Eqs. (\ref{2.4x}) and (\ref{2.4y}).

\noindent{\bf Energy:} Corresponding to the ``pulse lattice'' solution 
[Eq. (\ref{5.1}) with $G=0$] 
the energy is given by
\bea
&&\hat{E}=\frac{2D}{3m}\bigg ([(2-m)mB^2+(2m-1)A^2]E(m)
+(1-m)(A^2-2mB^2)K(m)] \bigg )\,, \nonumber \\
&&C=-\frac{1}{2}A^2D^2(1-m)-F^2[\alpha_1+3\beta_1 F^2-2H_z \rho_2 F]
+[-\gamma F^2 +\alpha_2  +\beta_2 (1-m)B^2]B^2\,.
\eea
For $m$ near one, the energy of this periodic solution can
be rewritten as the energy of the corresponding hyperbolic
(dark-dark) soliton solution [Eq. (\ref{2t})]  
plus the interaction energy. We find
\be
\hat{E}=E_{pulse}+E_{int}=D\left[\frac{2}{3}(A^2+B^2)
+\frac{(1-m)}{6}(3B^2-5A^2)+(1-m)(A^2-B^2)\ln\left(\frac{4}{\sqrt{1-m}} 
\right)\right] . 
\ee
Note that this solution also exists only when $\gamma < 0$, $4\beta_1 
\beta_2 < \gamma^2$.  Again, the interaction energy vanishes at $m=1$. 
It is amusing to note that the energy of the solution V is easily 
obtained from that of solution IV by simply interchanging $A$ and $B$.

\subsection{Solution VI}

Apart from the solutions which at $m=1$ reduce to the bright-bright and
dark-dark solutions, there are four solutions which at $m=1$ go over to
either bright-dark or dark-bright solutions, which we now discuss one by
one.

One such solution, kink-like in $\phi$ and pulse-like in $\psi$, is
\be\label{6.1}
\phi =F+A\sn[D(x+x_0),m]\,,~~\psi =G+B\cn[D(x+x_0),m]\,,
\ee
provided $G=0$ and the following seven coupled equations are satisfied
\be
2\alpha_1 F+4\beta_1 F^3+2\gamma FB^2-H_z\rho_1-3H_z \rho_2F^2
-H_z\rho_3B^2 =0\,,
\ee
\be
2\alpha_1 A+12\beta_1 F^2 A+2\gamma AB^2
-6H_z \rho_2 AF =-(1+m)AD^2\,,
\ee
\be
12\beta_1 F A^2-2\gamma FB^2-3H_z \rho_2A^2
+H_z\rho_3B^2 =0\,,
\ee
\be
2\beta_1 A^2-\gamma B^2=mD^2\,,
\ee
\be
2\alpha_2 B+2\gamma A^2B+2\gamma BF^2
-2H_z\rho_3 BF =(2m-1)BD^2\,,
\ee
\be
2\beta_2 B^2-\gamma A^2=-mD^2\,,
\ee
\be
(4\gamma F-2H_z\rho_3)AB=0\,.
\ee

In this case it turns out that the solution exists only if either
$2\beta_1 > \gamma > 2\beta_2$, $\gamma^2 >4\beta_1 \beta_2$ or
$2\beta_2 > \gamma > 2\beta_1$, $\gamma^2 < 4\beta_1 \beta_2$. 
We find that
\be\label{6.2}
A^2=\frac{mD^2[\gamma-2\beta_2]}{[\gamma^2-4\beta_1 \beta_2]}\,, ~~~ 
B^2=\frac{mD^2[2\beta_1-\gamma]}{[\gamma^2-4\beta_1 \beta_2]}\,,
\ee
where
\be
-(1+m)D^2=2\gamma B^2+\frac{2\rho_1 \gamma}{\rho_3}
-\frac{\rho_2 \rho_3 H_z^2}{2\gamma}\,,
~~F=\frac{\rho_3H_z}{2\gamma}\,,
\ee
and the three constraints are
\bea
&&\beta_1=\frac{\rho_2 \gamma}{2 \rho_3}\,,~~
\alpha_1=\frac{\rho_2 \rho_3 H_z^2}{2\gamma}+\frac{\rho_1
\gamma}{\rho_3}\,, \nonumber \\
&&(2m-1)D^2=2\alpha_2+2\gamma A^2-\frac{H_z^2
\rho_3^2}{2\gamma}\,.
\eea

\noindent{\bf Special case of $H_z=0$:}

With $H_z=0$, the solution is again given by Eq. (\ref{6.1}) but with 
$F=0=G$  where $A$ and $B$ are again as given by Eq. (\ref{6.2}) and 
furthermore, $\alpha_1,\alpha_2$ turn out to be negative, i.e. 
\be
\alpha_1=-\frac{(1+m)D^2}{2}-\gamma B^2\,,~
\alpha_2=-\frac{D^2}{2}-2\beta_2 B^2\,.
\ee

\noindent{\bf Special case of $\gamma^2 = 4\beta_1 \beta_2$}

One can show that the solution (\ref{6.1}) exists even in case $\gamma^2=
4\beta_1 \beta_2$. It turns out  such a solution
exists only if 
\be
2\beta_1=2\beta_2=\gamma\,,
\ee
and that in this case one cannot determine $A,B$. However, they must
satisfy the constraint
\be\label{1d}
A^2-B^2=\frac{mD^2}{\gamma}\,.
\ee
Further, one has
\be
\rho_2 = \rho_3\,,~~\alpha_1=\frac{\rho_3^2 H_z^2}{2\gamma}+
\frac{\rho_1 \gamma}{\rho_3}\,,~~
\alpha_2-\alpha_1 =\frac{mD^2}{2}-\frac{\rho_3^2 H_z^2}{2\gamma}\,.
\ee

\noindent{\bf Energy:} Corresponding to the mixed ``kink-pulse lattice'' 
solution [Eq. (\ref{6.1})] 
the energy is given by
\bea
&&\hat{E}=\frac{2D}{3m}\bigg ([(2-m)mB^2+(1+m)A^2]E(m)
-(1-m)(A^2+2B^2)K(m) \bigg )\,, \nonumber \\
&&C=-\frac{1}{2}A^2D^2-F^2[\alpha_1+3\beta_1 F^2-2H_z \rho_2 F]
+[-\gamma F^2 +\alpha_2  +\beta_2 B^2]B^2\,.
\eea
For $m$ near one, the energy of this periodic solution can
be rewritten as the energy of the corresponding hyperbolic
(bright-dark) soliton solution
\be\label{6t}
\phi =F+A\tanh[D(x+x_0)]\,,~~\psi =B\sech[D(x+x_0)]\,,
\ee
plus the interaction energy. We find
\be
\hat{E}=E_{soliton}+E_{int}=D\left[\frac{2}{3}(2A^2+B^2)
+\frac{(1-m)}{6}(2A^2+3B^2)-(1-m)B^2\ln\left(\frac{4}{\sqrt{1-m}} 
\right)\right]. 
\ee
Note that this solution also exists only when
either $2\beta_1 \ge \gamma \ge 2\beta_2$ and $\gamma^2 \ge 4\beta_1 \beta_2$
or $2\beta_2 \ge \gamma \ge 2\beta_1$ and $\gamma^2 \le 4\beta_1 \beta_2$.
The interaction energy vanishes at $m=1$. 
\subsection{Solution VII}

It is easy to show that another such (kink- and pulse-like) solution is
\be\label{7.1}
\phi =F+A\sn[D(x+x_0),m]\,,~~\psi =G+B\dn[D(x+x_0),m]\,,
\ee
provided $G=0$ and the following seven coupled equations are satisfied
\be
2\alpha_1 F+4\beta_1 F^3+2\gamma FB^2-H_z\rho_1-3H_z \rho_2F^2
-H_z\rho_3B^2 =0\,,
\ee
\be
2\alpha_1 A+12\beta_1 F^2 A+2\gamma AB^2
-6H_z \rho_2 AF =-(1+m)AD^2\,,
\ee
\be
12\beta_1 F A^2-2m\gamma FB^2-3H_z \rho_2A^2
+mH_z\rho_3B^2 =0\,,
\ee
\be
2\beta_1 A^2-m\gamma B^2=mD^2\,,
\ee
\be
2\alpha_2 B+(2/m)\gamma A^2B+2\gamma BF^2
-2H_z\rho_3 BF =(2-m)BD^2\,,
\ee
\be
2m\beta_2 B^2-\gamma A^2=-mD^2\,,
\ee
\be
4\gamma FAB-2H_z\rho_3AB=0\,.
\ee

In this case also it turns out that the solution exists only if either
$2\beta_1 > \gamma > 2\beta_2$, $\gamma^2 >4\beta_1 \beta_2$ or
$2\beta_2 > \gamma > 2\beta_1, \gamma^2 < 4\beta_1 \beta_2$. 
We find that
\be\label{7.2}
A^2=\frac{mD^2[\gamma-2\beta_2]}{[\gamma^2-4\beta_1 \beta_2]}\,, ~~~ 
B^2=\frac{D^2[2\beta_1-\gamma]}{[\gamma^2-4\beta_1 \beta_2]}\,,
\ee
where
\be\label{7.3}
-(1+m)D^2=2\gamma B^2+\frac{2\rho_1 \gamma}{\rho_3}
-\frac{\rho_2 \rho_3 H_z^2}{2 \gamma}\,,
~~F=\frac{\rho_3H_z}{2\gamma}\,,
\ee
and the three constraints are
\bea\label{7.4}
&&\beta_1=\frac{\rho_2 \gamma}{2 \rho_3}\,,~~
\alpha_1=\frac{\rho_2 \rho_3 H_z^2}{2\gamma}+\frac{\rho_1
\gamma}{\rho_3}\,, \nonumber \\
&&(2-m)D^2=2\alpha_2+(2/m)\gamma A^2-\frac{H_z^2
\rho_3^2}{2\gamma}\,.
\eea

\noindent{\bf Special case of $H_z=0$:}

With $H_z=0$, the solution is again given by Eq. (\ref{7.1}) but with 
$F=0=G$ where $A$ and $B$ are again as given by Eq. (\ref{7.2}) and 
furthermore, $\alpha_1,\alpha_2$ turn out to be negative, i.e. 
\be
\alpha_1=-\frac{(1+m)D^2}{2}-\gamma B^2\,,~
\alpha_2=-\frac{m D^2}{2}-2\beta_2 B^2\,.
\ee

\noindent{\bf m=1:}

In the special case of $m=1$ and $G=0,F \ne 0$, both solutions VI and 
VII reduce to a bright-dark type of solution as given by Eq. (\ref{6t}),
i.e.
\be\label{7.5}
\phi =F+A\tanh[D(x+x_0)]\,,~~\psi =B\sech[D(x+x_0)]\,,
\ee
with $A$, $B$ and $D$ given by
\be\label{7.6}
A^2=\frac{D^2[\gamma-2\beta_2]}{[\gamma^2-4\beta_1 \beta_2]}\,, ~~~ 
B^2=\frac{D^2[2\beta_1-\gamma]}{[\gamma^2-4\beta_1 \beta_2]}\,, ~~~ 
D^2=-\gamma B^2-\frac{\rho_1 \gamma}{\rho_3}
+\frac{\rho_2 \rho_3 H_z^2}{4 \gamma}\,,
\ee
while the other relations remain unchanged and are again given by Eqs.
(\ref{7.3}) and (\ref{7.4}).

\noindent{\bf Special case of $\gamma^2 = 4\beta_1 \beta_2$}

One can show that the solution (\ref{7.1}) exists even in case $\gamma^2=
4\beta_1 \beta_2$. It turns out  such a solution
exists only if 
\be
2\beta_1=2\beta_2=\gamma\,,
\ee
and that in this case one cannot determine $A,B$. However, they must
satisfy the constraint
\be\label{1e}
A^2-mB^2=\frac{mD^2}{\gamma}\,.
\ee
Further, one has
\be
\rho_2 = \rho_3\,,~~\alpha_1=\frac{\rho_3^2 H_z^2}{2\gamma}+
\frac{\rho_1 \gamma}{\rho_3}\,,~~
\alpha_2-\alpha_1 =\frac{D^2}{2}-\frac{\rho_3^2 H_z^2}{2\gamma}\,.
\ee

\noindent{\bf Energy:} Corresponding to the mixed ``kink-pulse lattice'' 
solution [Eq. (\ref{7.1})] 
the energy and the constant $C$ are given by
\bea
&&\hat{E}=\frac{2D}{3m}\bigg ([(2m-1)B^2+(1+m)A^2]E(m)
-(1-m)(A^2-B^2)K(m) \bigg )\,, \nonumber \\
\eea
\be 
C=-\frac{1}{2}A^2D^2-F^2[\alpha_1+3\beta_1 F^2-2H_z\rho_2 F] 
+ [-\gamma F^2+\alpha_2+\beta_2 B^2]B^2 . 
\ee 
For $m$ near one, the energy of this periodic solution can
be rewritten as the energy of the corresponding hyperbolic
(bright-dark) soliton solution [Eq. (\ref{6t})]   
plus the interaction energy. We find
\be
E=E_{soliton}+E_{int}=D\left[\frac{2}{3}(2A^2+B^2)
+\frac{(1-m)}{6}(2A^2-5B^2)+(1-m)B^2\ln\left(\frac{4}{\sqrt{1-m}} 
\right)\right] . 
\ee
Note that this solution also exists only when
either $2\beta_1 \ge \gamma \ge 2\beta_2$ and $\gamma^2 \ge 4\beta_1 \beta_2$
or $2\beta_2 \ge \gamma \ge 2\beta_1$ and $\gamma^2 \le 4\beta_1 \beta_2$.
The interaction energy vanishes for $m=1$. 

\subsection{Solution VIII}

Finally, we discuss two periodic solutions both of which at $m=1$ reduce  
to a dark-bright type of solution. The first such (pulse-like in $\phi$ 
and kink-like in $\psi$) solution is given by
\be\label{8.1}
\phi =F+A\cn[D(x+x_0),m]\,,~~\psi =G+B\sn[D(x+x_0),m]\,,
\ee
provided $G=0$ and the following seven coupled equations are satisfied
\be
2\alpha_1 F+4\beta_1 F^3+2\gamma FB^2-H_z\rho_1-3H_z \rho_2F^2
-H_z\rho_3B^2 =0\,,
\ee
\be
2\alpha_1 A+12\beta_1 F^2 A+2\gamma AB^2
-6H_z \rho_2 AF =(2m-1)AD^2\,,
\ee
\be
12\beta_1 F A^2-2\gamma FB^2-3H_z \rho_2A^2
+H_z\rho_3B^2 =0\,,
\ee
\be
2\beta_1 A^2-\gamma B^2=-mD^2\,,
\ee
\be
2\alpha_2 B+2\gamma A^2B+2\gamma BF^2
-2H_z\rho_3 BF =-(1+m)BD^2\,,
\ee
\be
2\beta_2 B^2-\gamma A^2=mD^2\,,
\ee
\be
4\gamma FAB-2H_z\rho_3AB=0\,.
\ee

In this case it turns out that the solution exists only if either
$2\beta_1 > \gamma > 2\beta_2$, $\gamma^2 <4\beta_1 \beta_2$ or
$2\beta_2 > \gamma > 2\beta_1$, $\gamma^2 > 4\beta_1 \beta_2$. 
We find that
\be\label{8.2}
A^2=\frac{mD^2[\gamma-2\beta_2]}{[4\beta_1 \beta_2-\gamma^2]}\,, ~~~ 
B^2=\frac{mD^2[2\beta_1-\gamma]}{[4\beta_1 \beta_2-\gamma^2]}\,,
\ee
where
\be\label{8.3}
(2m-1)D^2=2\gamma B^2+\frac{2\rho_1 \gamma}{\rho_3}
-\frac{\rho_2 \rho_3 H_z^2}{2\gamma}\,,
~~F=\frac{\rho_3H_z}{2\gamma}\,,
\ee
and the three constraints are
\bea\label{8.4}
&&\beta_1=\frac{\rho_2 \gamma}{2 \rho_3}\,,~~
\alpha_1=\frac{\rho_2 \rho_3 H_z^2}{2\gamma}+\frac{\rho_1
\gamma}{\rho_3}\,, \nonumber \\
&&-(1+m)D^2=2\alpha_2+2\gamma A^2-\frac{H_z^2 \rho_3^2}{2\gamma}\,.
\eea

\noindent{\bf Special case of $H_z=0$:}

When $H_z=0$, the solution is again given by Eq. (\ref{8.1}) but with 
$F=0=G$ where $A$ and $B$ are again as given by Eq. (\ref{8.2}) and 
furthermore, $\alpha_1$, $\alpha_2$ turn out to be negative, i.e. 
\be
\alpha_1=-\frac{ D^2}{2}-2\beta_1 A^2\,,~
\alpha_2=-\frac{(1+m)D^2}{2}-\gamma A^2\,.
\ee
It is worth pointing out that with $H_z=0$, the field equations are
completely symmetric between the two fields $\phi$ and $\psi$ and 
hence the solutions VI and VIII are identical in the limit $H_z=0$.

\noindent{\bf Special case of $\gamma^2 = 4\beta_1 \beta_2$}

One can show that the solution (\ref{8.1}) exists even in case $\gamma^2=
4\beta_1 \beta_2$. It turns out  such a solution
exists only if 
\be
2\beta_1=2\beta_2=\gamma\,,
\ee
and that in this case one cannot determine $A,B$. However, they must
satisfy the constraint
\be
B^2-A^2=\frac{mD^2}{\gamma}\,.
\ee
Further, one has
\be
\rho_2 = \rho_3\,,~~\alpha_1=\frac{\rho_3^2 H_z^2}{2\gamma}+
\frac{\rho_1 \gamma}{\rho_3}\,,~~
\alpha_1-\alpha_2 =\frac{mD^2}{2}+\frac{\rho_3^2 H_z^2}{2\gamma}\,.
\ee

\noindent{\bf Energy:} Corresponding to the mixed ``kink-pulse lattice'' 
solution [Eq. (\ref{8.1})] 
the energy and the constant $C$ are given by
\be
E-V_{min}=\frac{2D}{3m}\bigg ([(2m-1)A^2+(1+m)B^2]E(m)
-(1-m)(B^2-A^2)K(m) \bigg )\,, 
\ee
\be 
C=-\frac{1}{2}(1-m)A^2D^2-F^2[\alpha_1+3\beta_1 F^2-2H_z\rho_2 F] 
+ [-\gamma F^2 + \alpha_2 + \beta_2 B^2]B^2 . 
\ee 
For $m$ near one, the energy of this periodic solution can
be rewritten as the energy of the corresponding hyperbolic
(dark-bright) soliton solution
\be\label{8t}
\phi =F+A\sech[D(x+x_0)]\,,~~\psi =B\tanh[D(x+x_0)]\,,
\ee
plus the interaction energy. We find
\be
E=E_{soliton}+E_{int}=D\left[\frac{2}{3}(2B^2+A^2) +\frac{(1-m)}{6} 
(2B^2-5A^2)+(1-m)A^2\ln\left(\frac{4}{\sqrt{1-m}}\right)\right] . 
\ee
Note that this solution also exists only when either $2\beta_1 \ge \gamma 
\ge 2\beta_2$ and $\gamma^2 \le 4\beta_1 \beta_2$ or $2\beta_2 \ge \gamma \ge 
2\beta_1$ and $\gamma^2 \ge 4\beta_1 \beta_2$.  Again, the interaction 
energy vanishes at $m=1$.  It is amusing to note that the energy of the 
solution VIII is easily obtained from that of solution VII by simply 
interchanging $A$ and $B$.

\subsection{Solution IX}

Another solution which at $m=1$ reduces to a dark-bright type of soliton 
solution is given by
\be\label{9.1}
\phi =F+A\dn[D(x+x_0),m]\,,~~\psi =G+B\sn[D(x+x_0),m]\,,
\ee
provided $G=0$ and the following seven coupled equations are satisfied
\be
2\alpha_1 F+4\beta_1 F^3+(2/m)\gamma FB^2-H_z\rho_1-3H_z \rho_2F^2
-\frac{H_z\rho_3B^2}{m} =0\,,
\ee
\be
2\alpha_1 A+12\beta_1 F^2 A+(2/m)\gamma AB^2
-6H_z \rho_2 AF =(2-m)AD^2\,,
\ee
\be
12\beta_1 F A^2-(2/m)\gamma FB^2-3H_z \rho_2A^2
+\frac{H_z\rho_3B^2}{m} =0\,,
\ee
\be
2m\beta_1 A^2-\gamma B^2=-mD^2\,,
\ee
\be
2\alpha_2 B+2\gamma A^2B+2\gamma BF^2
-2H_z\rho_3 BF =-(1+m)BD^2\,,
\ee
\be
2\beta_2 B^2-m\gamma A^2=mD^2\,,
\ee
\be
4\gamma FAB-2H_z\rho_3AB=0\,.
\ee

In this case it turns out that the solution exists only if either
$2\beta_1 > \gamma > 2\beta_2$, $\gamma^2 <4\beta_1 \beta_2$ or
$2\beta_2 > \gamma > 2\beta_1$, $\gamma^2 > 4\beta_1 \beta_2$. 
We find that
\be\label{9.2}
A^2=\frac{D^2[\gamma-2\beta_2]}{[4\beta_1 \beta_2-\gamma^2]}\,, ~~~ 
B^2=\frac{mD^2[2\beta_1-\gamma]}{[4\beta_1 \beta_2-\gamma^2]}\,,
\ee
where
\be\label{9.3}
(2-m)D^2=(2/m)\gamma B^2+\frac{2\rho_1 \gamma}{\rho_3}
-\frac{\rho_2 \rho_3 H_z^2}{2\gamma}\,,
~~F=\frac{\rho_3H_z}{2\gamma}\,,
\ee
and the three constraints are
\bea\label{9.4}
&&\beta_1=\frac{\rho_2 \gamma}{2 \rho_3}\,,~~
\alpha_1=\frac{\rho_2 \rho_3 H_z^2}{2\gamma}+\frac{\rho_1
\gamma}{\rho_3}\,, \nonumber \\
&&-(1+m)D^2=2\alpha_2+2\gamma A^2-\frac{H_z^2 \rho_3^2}{2\gamma}\,.
\eea

\noindent{\bf Special case of $H_z=0$:}

When $H_z=0$, the solution is again given by Eq. (\ref{9.1}) but with 
$F=0=G$  where $A$ and $B$ are again as given by Eq. (\ref{9.2}) and 
furthermore, $\alpha_1,\alpha_2$ turn out to be negative, i.e. 
\be
\alpha_1=-\frac{m D^2}{2}-2\beta_1 A^2\,,~
\alpha_2=-\frac{(1+m)D^2}{2}-\gamma A^2\,.
\ee
It is worth pointing out that with $H_z=0$, the field equations are
completely symmetric between the two fields $\phi$ and $\psi$ and hence
the solutions VII and IX are identical in the limit $H_z=0$.

\noindent{\bf m=1:}

In the special case of $m=1$ and $G=0,F \ne 0$, both solutions VIII 
and IX reduce to a dark-bright type of solution (\ref{8t}), i.e.
\be\label{9.5}
\phi =F+A\sech[D(x+x_0)]\,,~~\psi =B\tanh[D(x+x_0)]\,,
\ee
with $A$, $B$ and $D$ given by
\be\label{9.6}
A^2=\frac{D^2[\gamma-2\beta_2]}{[4\beta_1 \beta_2-\gamma^2]}\,, ~~~ 
B^2=\frac{D^2[2\beta_1-\gamma]}{[4\beta_1 \beta_2-\gamma^2]}\,, ~~~ 
D^2=2\gamma B^2+\frac{2\rho_1 \gamma}{\rho_3}
-\frac{\rho_2 \rho_3 H_z^2}{2\gamma}\,,
\ee
while the other relations remain unchanged and are given by Eqs.
(\ref{8.3}) and (\ref{8.4}).

\noindent{\bf Special case of $\gamma^2 = 4\beta_1 \beta_2$}

One can show that the solution (\ref{9.1}) exists even in case $\gamma^2=
4\beta_1 \beta_2$. It turns out  such a solution
exists only if 
\be
2\beta_1=2\beta_2=\gamma\,,
\ee
and that in this case one cannot determine $A,B$. However, they must
satisfy the constraint
\be
B^2-mA^2=\frac{mD^2}{\gamma}\,.
\ee
Further, one has
\be
\rho_2 = \rho_3\,,~~\alpha_1=\frac{\rho_3^2 H_z^2}{2\gamma}+
\frac{\rho_1 \gamma}{\rho_3}\,,~~
\alpha_1-\alpha_2 =\frac{D^2}{2}+\frac{\rho_3^2 H_z^2}{2\gamma}\,.
\ee

\noindent{\bf Energy:} Corresponding to the mixed ``pulse-kink lattice'' 
solution [Eq. (\ref{9.1})] 
the energy and the constant $C$ are given by
\be
E-V_{min}=\frac{2D}{3m}\bigg ([(2-m)mA^2+(1+m)B^2]E(m)
-(1-m)(B^2+2A^2)K(m) \bigg )\,, 
\ee
\be
C=\frac{1}{2}(1-m)A^2D^2-F^2[\alpha_1+3\beta_1 F^2-2H_z\rho_2 F]
+ [-m\gamma F^2 + m\alpha_2 + \beta_2 B^2]\frac{B^2}{m^2} .
\ee
For $m$ near one, the energy of this periodic solution can
be rewritten as the energy of the corresponding hyperbolic
(dark-bright) soliton solution [Eq. (\ref{8t})]  
plus the interaction energy. We find
\be
E=E_{soliton}+E_{int}=D\left[\frac{2}{3}(2B^2+A^2) +\frac{(1-m)}{6} 
(2B^2+3A^2)-(1-m)A^2\ln\left(\frac{4}{\sqrt{1-m}}\right)\right] . 
\ee

Note that this solution also exists only when either $2\beta_1 \ge \gamma 
\ge 2\beta_2$ and $\gamma^2 \le 4\beta_1 \beta_2$ or $2\beta_2 \ge \gamma
\ge 2\beta_1$ and $\gamma^2 \ge 4\beta_1 \beta_2$.  The interaction energy 
vanishes at $m=1$.  It is amusing to note that the energy of the solution 
IX is easily obtained from that of solution VI by simply interchanging 
$A$ and $B$.

Summarizing, we have obtained nine periodic solutions in terms of Jacobi
elliptic functions, in the case of coupled $\phi^4$ field theory with 
biquadratic coupling and an external magnetic field. This was possible
because the magnetic field interaction is not symmetric between the two
fields $\phi$ and $\psi$. In the special case when the modulus parameter
$m$ of the Jacobi elliptic function is one, these nine solutions reduce to 
four different soliton solutions valid on the full line and expressed in terms 
of hyperbolic functions. Note, however, that in case the external field
$H_z=0$, instead of nine, we only obtain six distinct periodic solutions 
(of which three are previously known \cite{liu}), which in the limit $m=1$, 
give three distinct soliton solutions. 

It is worth emphasizing the restrictions on the various parameters in the 
case of the nine solutions.  For example, in the case of $\sn-\sn$ solution 
(with $G=0$), $2\beta_1 \ge \gamma$, $2\beta_2 \ge \gamma$. Further, in the 
special case of $H_z=0=F=G$, one can show that $\alpha_1<0$, $\alpha_2<0$.
On the other hand, in the case of $\cn-\cn$, $\dn-\dn$, $\cn-\dn$ and 
$\dn-\cn$ solutions (with $G=0$), $\gamma < 0$ and further $\gamma^2 >4 
\beta_1 \beta_2$.
In the special case of $H_z=0=F=G$, in addition one finds that (i) for
$\cn-\cn$ case $\alpha_1$, $\alpha_2 > (<)$ 0 provided $m > (<)$ 1/2 (ii) 
$\alpha_1>0$, $\alpha_2>0$ for $\dn-\dn$ solution (iii) for $\dn-\cn$
solution, $\alpha_2 >0$ (iv) $\alpha_1 <0$ for $\cn-\dn$ solution.
Instead, for $\sn-\cn$ as well as $\sn-\dn$ solutions, either $2\beta_1 \ge 
\gamma \ge 2\beta_2$ and $\gamma^2 \ge 4\beta_1\beta_2$ or $2\beta_2 \ge 
\gamma \ge 2\beta_1$ and $\gamma^2 \le 4\beta_2\beta_4$. Finally,
in the $\cn-\sn$ and $\dn-\sn$  cases,
either $2\beta_1 \ge \gamma \ge 2\beta_2$ and
$\gamma^2 \le 4\beta_1\beta_2$ or $2\beta_2 \ge \gamma \ge 2\beta_1$ and 
$\gamma^2 \ge 4\beta_1\beta_2$. If in addition $H_z=0=F=G$, then for all 
the four solutions (i.e. $\sn-\cn$, $\cn-\sn$, $\sn-\dn$, $\dn-\sn$), 
$\alpha_1<0$ and $\alpha_2 <0$.

\section{Solutions with bilinear coupling}

Several years ago, a coupled $\phi^4$ model was considered in the context 
of a surface phase transition with hydration forces \cite{jcp}
which is similar to the one considered in the last section except that
there was no external magnetic field $H_z$ and instead of a biquadratic
coupling there was a bilinear coupling between the two fields. The 
purpose of this section is to obtain a bright-bright soliton solution 
of that model. 

The potential (i.e. free energy) is given by 
\be
V=\alpha_1 \phi^2 +\beta_1 \phi^4 +\alpha_2 \psi^2 +\beta_2 \psi^4 
+\delta_1 (\phi - \psi)^2 +\delta_2 (\phi+\psi)^2\,, 
\ee
with $\delta_1$, $\delta_2$ being the coupling parameters between the two
fields. The (static) equations of motion which follow from here are
\be\label{10t}
\frac{d^2 \phi}{dx^2}=2\alpha_1 \phi +4\beta_1 \phi^3 
+2\delta_1 (\phi - \psi)+2\delta_2 (\phi + \psi)\,, 
\ee
\be\label{10tt}
\frac{d^2 \psi}{dx^2}=2\alpha_2 \psi +4\beta_2 \psi^3 
-2\delta_1 (\phi - \psi)+2\delta_2 (\phi + \psi)\,. 
\ee

It is not difficult to show that this pair of coupled field equations 
admits the ``kink-kink" type periodic solution
\be\label{10.1}
\phi =A\sn[D(x+x_0),m]\,,~~\psi =B\sn[D(x+x_0),m]\,,
\ee
provided 
\bea\label{3.3xx}
&&mD^2=2\beta_1 A^2 = 2 \beta_2 B^2\,, \nonumber \\
&&A^2=\frac{-m}{(1+m)\beta_1}\left[\alpha_1+\delta_1+\delta_2 
+(\delta_2-\delta_1)\sqrt{\frac{\beta_1}{\beta_2}}\right]\,,
\eea
and furthermore the parameters $\alpha_1,\alpha_2,\beta_1,\beta_2, 
\delta_1, \delta_2$ satisfy the constraint
\be\label{3.3x}
\alpha_2 - \alpha_1=(\delta_1 -\delta_2)\left[\sqrt{\frac{\beta_2} 
{\beta_1}}-\sqrt{\frac{\beta_1}{\beta_2}}\right]\,.
\ee
Since $A^2>0$, the relation (\ref{3.3xx}) gives us a strong constraint 
on some of the parameters. The energy $\hat{E}$ and the constant $C$ 
corresponding to the periodic solution [Eq. (\ref{10.1})] are 
\bea
&&\hat{E}=\frac{2(A^2+B^2)D}{3m}[(1+m)E(m)-(1-m)K(m)]\,, \nonumber \\
&&C=-\frac{1}{2}(A^2+B^2)D^2\,.
\eea

In the limit $m=1$, this solution reduces to the bright-bright soliton solution
\be\label{10.2}
\phi =A\tanh[D(x+x_0),m]\,,~~\psi =B\tanh[D(x+x_0)]\,,
\ee
provided 
\bea
&&D^2=2\beta_1 A^2 = 2 \beta_2 B^2\,, \nonumber \\
&&A^2=\frac{-1}{2\beta_1}\left[\alpha_1+\delta_1+\delta_2+(\delta_2 
-\delta_1)\sqrt{\frac{\beta_1}{\beta_2}}\right]\,,
\eea
while the relation (\ref{3.3x}) remains unchanged. 
For $m$ near one, the energy of the periodic solution can be rewritten 
as the energy of the corresponding hyperbolic (bright-bright) soliton 
solution [Eq. (\ref{10.2})] plus the interaction energy. We find
\be
\hat{E}=E_{kink}+E_{int}=(A^2+B^2)D \left[\frac{4}{3}+\frac{(1-m)}{3}\right]\,.
\ee

The interaction energy vanishes at $m=1$.  In view of the requirement 
$\beta_1$, $\beta_2 > 0$ arising from stability, we are unable to find any 
other solution to this coupled set of equations with a bilinear coupling.

\section{Solutions of discrete coupled $\phi^4$-type equations with
biquadratic coupling}

Discrete coupled $\phi^4$ models arise in the context of structural 
transitions on a lattice, collective proton dynamics in ice \cite{ice}, 
etc.  The purpose of this section is to give an exhaustive list of 
solutions to the discrete coupled $\phi^4$-type equations with 
biquadratic coupling (but in the absence of an external magnetic field 
$H_z$). In the next section, we shall obtain a solution of the discrete 
coupled $\phi^4$-type equations with bilinear coupling.

We start from the coupled static field equations (\ref{2}) and (\ref{3}). 
The discrete analog of these field equations, for $H_z=0$ has the form

\be\label{4}
   \frac{1}{h^2} (\phi_{n+1}+\phi_{n-1}-2\phi_n)
   -2\alpha_1\phi_n-2[2\beta_1\phi_n^2+\gamma \psi_n^2]\phi_n=0\,, 
\ee
\be\label{5}
   \frac{1}{h^2} (\psi_{n+1}+\psi_{n-1}-2\psi_n)
   -2\alpha_2\psi_n-2[2\beta_2\psi_n^2+\gamma \phi_n^2]\psi_n=0\,.
\ee
We are unable to find any solution to this coupled set of field
equations.  However, as in the Ablowitz-Ladik discretization of the 
discrete nonlinear Schr\"odinger equation \cite{al}, if we replace
$\phi_n$ and $\psi_n$ in the last term in 
Eqs. (\ref{4}) and (\ref{5}) by their average,
then we can find exact solutions to this coupled system. In particular,
instead of Eqs. (\ref{4}) and (\ref{5}), we consider the discretized
equations
\be\label{6}
   \frac{1}{h^2} (\phi_{n+1}+\phi_{n-1}-2\phi_n)
   -2\alpha_1\phi_n-[2\beta_1\phi_n^2+\gamma \psi_n^2]
   [\phi_{n+1}+\phi_{n-1}]=0\,, 
\ee
\be\label{7}
   \frac{1}{h^2} (\psi_{n+1}+\psi_{n-1}-2\psi_n)
   -2\alpha_2\psi_n-[2\beta_2\psi_n^2+\gamma \phi_n^2]
   [\psi_{n+1}+\psi_{n-1}]=0\,.
\ee
Note that single solitons and their stability in coupled Ablowitz-Ladik 
chains have been studied previously \cite{yang}.  We now show that this 
modified set of coupled discrete equations has six different periodic 
solutions which in the limit $m=1$ reduce to the bright-bright, 
bright-dark and dark-dark soliton solutions. In all the solutions, we
shall see that the static kink can be placed anywhere with respect to
the lattice. Hence we suspect that in all these cases, there may be
an absence of the Peierls-Nabarro barrier \cite{PN1,PN2,pgk}, which is 
the energy cost associated with moving a localized solution such as a 
soliton by a half lattice constant on a discrete lattice. It would be 
nice if one can demonstrate this explicitly.

\subsection{Solution I}

It is easy to show that the field Eqs. (\ref{6}) and (\ref{7}) admit
the kink-kink type solution
\be\label{7a}
\phi_n = A~ \sn[hD(n+x_0),m]\,, ~~~ 
\psi_n = B~ \sn[hD(n+x_0),m]\,,
\ee
provided 
\bea
&&A^2=\frac{m(2\beta_2-\gamma)\sn^2(hD,m)}{h^2(4\beta_1
\beta_2-\gamma^2)}\,, ~~~
B^2=\frac{m(2\beta_1-\gamma)\sn^2(hD,m)}{h^2(4\beta_1
\beta_2-\gamma^2)}\,, \nonumber \\
&&\alpha_1=\alpha_2=-\frac{1}{h^2}[1-\cn(hD,m)\dn(hD,m)],   
\eea
where $h$ is the lattice spacing.  Thus, note that as in the continuum 
case, this solution exists provided $2\beta_1>\gamma$, $2\beta_2 >\gamma$, 
$\alpha_1 <0$, $\alpha_2 <0$. It is interesting to note that the solutions 
to both the discrete and the continuum model exist under the same set 
of conditions.

\noindent{\bf Continuum Limit:} It is instructive to consider the continuum 
limit $h \rightarrow 0$ and show that the above solution smoothly goes  
over to the corresponding continuum solution. In particular, on using the 
fact that as $h \rightarrow 0$
\be\label{4xy}
\sn(hD,m) \rightarrow hD\,,~ \cn^2(hD,m) \rightarrow 1-h^2D^2\,,
~\dn^2(hD,m) \rightarrow 1-mh^2D^2\,,
\ee
it readily follows that the above solution indeed reduces to the
corresponding continuum solution [Eq. (7)] obtained in Sec. II 
(when $H_z=F=G=0$), i.e.
\be
A^2=\frac{m(2\beta_2-\gamma)D^2}{(4\beta_1 \beta_2 -\gamma^2)}\,, ~~~ 
B^2=\frac{m(2\beta_1-\gamma)D^2}{(4\beta_1 \beta_2 -\gamma^2)}\,, ~~~ 
\alpha_1=\alpha_2=-\frac{(1+m)D^2}{2}\,.
\ee
In fact we shall see that all the six solutions of this coupled discrete model
smoothly go over to the corresponding continuum solutions obtained in 
Sec. II in the limit $h \rightarrow 0$. 

In the limit $m=1$, the periodic solution (\ref{7a}) reduces to the 
bright-bright soliton solution
\be
\phi_n = A~ \tanh[hD(n+x_0)]\,, ~~~~ 
\psi_n = B~ \tanh[hD(n+x_0)]\,.
\ee

\noindent{\bf Special case of $\gamma^2 = 4\beta_1 \beta_2$}

One can show that the solution (\ref{7a}) exists even in case $\gamma^2=
4\beta_1 \beta_2$. It turns out that such a solution
exists only if 
\be
2\beta_1=2\beta_2=\gamma\,,
\ee
and that in this case one cannot determine $A,B$. However, they must
satisfy the constraint
\be
A^2+B^2=\frac{m\sn^2(hD,m)}{h^2\gamma}\,.
\ee
In the continuum limit $h \rightarrow 0$, as expected, this reduces to the
constraint equation (\ref{1c}) obtained in Sec. II.

\subsection{Solution II}

It is easily shown that a kink-pulse type solution 
\be\label{7b}
\phi_n = A~ \sn[hD(n+x_0),m]\,, ~~~ 
\psi_n = B~ \cn[hD(n+x_0),m]\,,
\ee
is an exact solution to the field Eqs. (\ref{6}) and (\ref{7}) provided 
\be
A^2=\frac{m(\gamma-2\beta_2)\sn^2(hD,m)}{h^2(\gamma^2-4\beta_1 
\beta_2)}\,, ~~~
B^2=\frac{m(2\beta_1-\gamma)\sn^2(hD,m)}{h^2(\gamma^2-4\beta_1
\beta_2)}\,.
\ee
Furthermore,
\be
-\alpha_1=\frac{1}{h^2}+\frac{\cn(hD,m)[2m\beta_1(\gamma-2\beta_2) \sn^2(hD,m)
-(\gamma^2-4\beta_1 \beta_2)]}{h^2(\gamma^2-4\beta_1 \beta_2)
\dn(hD,m)}\,,
\ee
\be
-\alpha_2=\frac{1}{h^2}+\frac{\cn(hD,m)[m\gamma(\gamma-2\beta_2)\sn^2(hD,m)
-(\gamma^2-4\beta_1 \beta_2)]}
{h^2(\gamma^2-4\beta_1 \beta_2)\dn^2(hD,m)}\,. 
\ee
Again, as in the continuum case either $2\beta_1>\gamma>2\beta_2$,
$\gamma^2> 4\beta_1 \beta_2$ or $2\beta_2>\gamma>2\beta_1$,
$4\beta_1 \beta_2>\gamma^2$. 
It is, however, not clear here if $\alpha_1$ and $\alpha_2$ 
have a definite sign. However, in the limit $m=1$, as in the 
continuum case, one finds that $\alpha_1 <0$, $\alpha_2 < 0$.

It is readily checked that in the continuum limit this solution smoothly
goes over to the corresponding continuum solution, Eqs. (104) and 
(\ref{6.2}).  Furthermore, the corresponding bright-dark solution is 
easily obtained in the limit $m=1$. 

\noindent{\bf Special case of $\gamma^2 = 4\beta_1 \beta_2$}

One can show that the solution (\ref{7b}) exists even in case $\gamma^2=
4\beta_1 \beta_2$. It turns out  such a solution
exists only if 
\be
2\beta_1=2\beta_2=\gamma\,,
\ee
and that in this case one cannot determine $A,B$. However, they must
satisfy the constraint
\be
A^2-B^2\dn^2(hD,m)=\frac{m\sn^2(hD,m)}{h^2\gamma}\,.
\ee
In the continuum limit $h \rightarrow 0$, as expected, this reduces to the
constraint equation (\ref{1d}) obtained in Sec. II.

\subsection{Solution III}

Yet another kink-pulse type solution is given by
\be
\phi_n = A~ \sn[hD(n+x_0),m]\,, ~~~ 
\psi_n = B~ \dn[hD(n+x_0),m]\,,
\ee
provided 
\be
A^2=\frac{m(\gamma-2\beta_2)\sn^2(hD,m)}{h^2(\gamma^2-4\beta_1 \beta_2)}\,, 
~~~
B^2=\frac{(2\beta_1-\gamma)\sn^2(hD,m)}{h^2(\gamma^2-4\beta_1
\beta_2)}\,.
\ee
Furthermore,
\be
-\alpha_1=\frac{1}{h^2}+\frac{\dn(hD,m)[2\beta_1(\gamma-2\beta_2)\sn^2(hD,m)
-(\gamma^2-4\beta_1 \beta_2)]}
{h^2(\gamma^2-4\beta_1 \beta_2)\cn(hD,m)}\,,
\ee
\be
-\alpha_2=\frac{1}{h^2}+\frac{\dn(hD,m)[\gamma(\gamma-2\beta_2)\sn^2(hD,m)
-(\gamma^2-4\beta_1 \beta_2)]}
{h^2(\gamma^2-4\beta_1 \beta_2)\cn^2(hD,m)}\,. 
\ee

\noindent{\bf Special case of $\gamma^2 = 4\beta_1 \beta_2$}

One can show that the solution (208) exists even in case $\gamma^2=
4\beta_1 \beta_2$. It turns out  such a solution
exists only if 
\be
2\beta_1=2\beta_2=\gamma\,,
\ee
and that in this case one cannot determine $A,B$. However, they must
satisfy the constraint
\be
A^2-mB^2\cn^2(hD,m)=\frac{m\sn^2(hD,m)}{h^2\gamma}\,.
\ee
In the continuum limit $h \rightarrow 0$, as expected, this reduces to the
constraint equation (\ref{1e}) obtained in Sec. II.

Note that in the $m=1$ limit the solutions II and III reduce to the 
same bright-dark soliton solution.  In addition, as in the continuum 
case, this solution exists if 
either $2\beta_1 \ge \gamma \ge 2\beta_2$, $\gamma^2 \ge 4\beta_1\beta_2$ 
or $2\beta_2 \ge \gamma \ge 2\beta_1$, $4\beta_1 \beta_2 \ge \gamma^2$.
It is, however, not clear here if $\alpha_1$ and $\alpha_2$ 
have a definite sign. However, in the limit $m=1$, as in the 
continuum case, one finds that $\alpha_1 <0$ and $\alpha_2 < 0$.

\subsection{Solution IV}

Finally, we present three periodic solutions, all of which in the limit 
$m=1$ reduce to the dark-dark soliton solution.  One of the pulse-pulse 
type periodic solution is given by
\be
\phi_n = A~ \cn[hD(n+x_0),m]\,, ~~~ 
\psi_n = B~ \cn[hD(n+x_0),m]\,,
\ee
provided as in the continuum case, 
$\gamma<0$ and $\gamma^2 >4 \beta_1 \beta_2$. We find
\bea
&&A^2=\frac{2m(\beta_2+|\gamma|)\sn^2(hD,m)}
{h^2(\gamma^2-4\beta_1 4\beta_2)\dn^2(hD,m)}\,, ~~~
B^2=\frac{2m(\beta_1+|\gamma|)\sn^2(hD,m)}
{h^2(\gamma^2-4\beta_1 4\beta_2)\dn^2(hD,m)}\,, \nonumber \\
&&\alpha_1=\alpha_2=-\frac{1}{h^2}\left[1-\frac{\cn(hD,m)}{\dn^2(hD,m)} 
\right]\,.
\eea
Using Eq. (\ref{4xy}) it is easily shown that as in the continuum
case, $\alpha_1$, $\alpha_2 > (<)$ 0 provided $m > (<)$ 1/2.  This 
solution is equivalent to the continuum solution, Eqs. (33) and (42).

\subsection{Solution V}

Another pulse-pulse type solution is given by
\be
\phi_n = A~ \dn[hD(n+x_0),m]\,, ~~~ 
\psi_n = B~ \dn[hD(n+x_0),m]\,,
\ee
provided as in the continuum case, 
$\gamma<0$ and $\gamma^2 >4 \beta_1 \beta_2$. We find
\bea
&&A^2=\frac{2(\beta_2+|\gamma|)\sn^2(hD,m)}
{h^2(\gamma^2-4\beta_1 4\beta_2)\cn^2(hD,m)}\,, ~~~
B^2=\frac{2(\beta_1+|\gamma|)\sn^2(hD,m)}
{h^2(\gamma^2-4\beta_1 4\beta_2)\cn^2(hD,m)}\,, \nonumber \\
&&\alpha_1=\alpha_2=-\frac{1}{h^2}\left[1-\frac{\dn(hD,m)}{\cn^2(hD,m)} 
\right]\,.
\eea
Using Eq. (\ref{4xy}) it is easily shown that as in the continuum
case, $\alpha_1$, $\alpha_2 > 0$. This solution is equivalent to the 
continuum solution, Eqs. (33) and (62).

\subsection{Solution VI}

Yet another pulse-pulse type solution is
\be
\phi_n = A~ \dn[hD(n+x_0),m]\,, ~~~ 
\psi_n = B~ \cn[hD(n+x_0),m]\,,
\ee
provided as in the continuum case, 
$\gamma<0$ and $\gamma^2 >4 \beta_1 \beta_2$. We find
\be
A^2=\frac{(2\beta_2+|\gamma|)\sn^2(hD,m)}
{h^2(\gamma^2-4\beta_1 \beta_2)\cn^2(hD,m)}\,, ~~~
B^2=\frac{m(2\beta_1+|\gamma|)\sn^2(hD,m)}
{h^2(\gamma^2-4\beta_1 \beta_2)\dn^2(hD,m)}\,. 
\ee
Furthermore,
\be
-\alpha_1=\frac{1}{h^2}+\frac{[2\beta_1(2\beta_2+|\gamma|)\dn^2(hD,m) 
-\gamma(2\beta_1+|\gamma|)\cn^2(hD,m))]}{h^2(\gamma^2-4\beta_1
\beta_2)\dn(hD,m)\cn^2(hD,m)}\,,
\ee
\be
-\alpha_2=\frac{1}{h^2}+\frac{[2\beta_2(2\beta_1+|\gamma|)\cn^2(hD,m) 
-|\gamma|(|\gamma|+2\beta_2)\dn^2(hD,m)]}{h^2(\gamma^2-4\beta_1
\beta_2)\dn^2(hD,m)\cn(hD,m)}\,.  
\ee
It is not clear if in general $\alpha_1$ and $\alpha_2$ have a definite 
sign. However, it is easily checked that at $m=1$, $\alpha_1$, $\alpha_2>0$.
This solution is equivalent to the continuum solution, Eqs. (72) and (80). 

Note that the last three (i.e. IV,V,VI) solutions reduce to the 
(same) dark-dark soliton solution in the limit $m=1$.  Since we do not 
know the Hamiltonian corresponding to Eqs. (193) and (194), we are unable 
to find the energy and soliton interaction explicitly for any of the above 
discrete solutions.  Similarly, for $H_z\ne0$ we have not succeeded in 
finding exact solutions.   
 
\section{Solutions of discrete coupled $\phi^4$-type equations with
bilinear coupling}

Coupled lattice chains, with a bilinear coupling, undergoing a second 
order structural phase transition can represent this case.  We start from 
the coupled static field Eqs. (\ref{10t}) and (\ref{10tt}).  The discrete 
analog of these field equations has the form
\be\label{10s}
   \frac{1}{h^2} (\phi_{n+1}+\phi_{n-1}-2\phi_n)
   -2[\alpha_1+\delta_1+\delta_2]\phi_n
   +2[\delta_1-\delta_2]\psi_n-4\beta_1 \phi_n^3=0\,, 
\ee
\be\label{10ss}
   \frac{1}{h^2} (\psi_{n+1}+\psi_{n-1}-2\psi_n)
   -2[\alpha_2+\delta_1+\delta_2]\psi_n
   +2[\delta_1-\delta_2]\phi_n-4\beta_2 \psi_n^3=0\,.  
\ee
We are unable to find any solution to this coupled set of field
equations.  However, as in the Ablowitz-Ladik discretization of the 
discrete nonlinear Schr\"odinger equation \cite{al}, if we replace
$\phi_n$ and $\psi_n$ in the last term in Eqs. (\ref{10s}) and 
(\ref{10ss}) by their average, then we can find exact solutions to 
this coupled system. In particular, instead of Eqs. (\ref{10s}) and 
(\ref{10ss}), we consider the discretized equations
\be\label{10u}
   \frac{1}{h^2} (\phi_{n+1}+\phi_{n-1}-2\phi_n)
   -2[\alpha_1+\delta_1+\delta_2]\phi_n
   +2[\delta_1-\delta_2]\psi_n-2\beta_1 \phi_n^2 
   [\phi_{n+1}+\phi_{n-1}]=0\,, 
\ee
\be\label{10uu}
   \frac{1}{h^2} (\psi_{n+1}+\psi_{n-1}-2\psi_n)
   -2[\alpha_2+\delta_1+\delta_2]\psi_n
   +2[\delta_1-\delta_2]\phi_n-2\beta_2 \psi_n^2 
   [\psi_{n+1}+\psi_{n-1}]=0\,.
\ee
It is easy to show that the field Eqs. (\ref{10u}) and (\ref{10uu}) admit
the kink-kink type solution
\be\label{10v}
\phi_n = A~ \sn[hD(n+x_0),m]\,, ~~~ 
\psi_n = B~ \sn[hD(n+x_0),m]\,,
\ee
provided
\bea\label{4.3xx}
&&\frac{m\sn^2(hD,m)}{h^2}=2\beta_1 A^2 = 2 \beta_2 B^2\,, \nonumber \\
&&A^2=\frac{-m\sn^2(hD,m)}{2\beta_1[1-\cn(hD,m)\dn(hD,m)]}
\left[\alpha_1+\delta_1+\delta_2 
+(\delta_2-\delta_1)\sqrt{\frac{\beta_1}{\beta_2}}\right]\,,
\eea
and furthermore the parameters $\alpha_1,\alpha_2,\beta_1,\beta_2, 
\delta_1, \delta_2$ satisfy the constraint
\be\label{4.3x}
\alpha_2 - \alpha_1=(\delta_1 -\delta_2)\left[\sqrt{\frac{\beta_2} 
{\beta_1}}-\sqrt{\frac{\beta_1}{\beta_2}}\right]\,.
\ee
As expected, in the continuum limit of $h \rightarrow 0$, this solution
smoothly goes over to the corresponding continuum solution, Eqs. (184), 
(185) and (186), obtained in Sec. III.  Since we do not know the 
Hamiltonian corresponding to Eqs. (224) and (225), we are unable to 
find the energy and soliton interation explcitly for this discrete 
solution.  Similarly, for $H_z\ne0$ we have not succeeded in finding 
an exact solution.  

\section{Conclusion} 

We have systematically provided an exhaustive set of exact periodic 
domain wall solutions for a coupled $\phi^4$ model with and without 
an external field, and for both bilinear and biquadratic couplings. 
Only a bright-bright solution could be obtained for the bilinear 
case.  For both the biquadratic and bilinear couplings the corresponding 
discrete case was also considered--with an Ablowitz-Ladik like 
modification of the coupled discrete equations-- and we obtained several 
exact solutions. For the solutions of the discrete model, the calculation 
of the Peierls-Nabarro barrier \cite{PN1,PN2,pgk} and soliton scattering 
\cite{krss,ckms} remain topics of further study.  Similarly, scattering 
of solitons in the coupled $\phi^4$ continuum and discrete models with 
either the biquadratic or bilinear coupling is an interesting open issue.  
To this end, the static solutions presented here need to be boosted with  
a certain velocity.  

It would be instructive to explore whether the nine different solutions 
reported in Sec. 2 (or the six solutions in Sec. 4) are completely 
disjoint or if there are any possible bifurcations linking them via, for 
instance, analytical continuation.  We have not tried to carry out an 
explicit stability analysis of various periodic solutions.  However, 
the energy calculations and interaction energy between solitons (for 
$m\sim1$) in the case of both the biquadratic and bilinear couplings  
could provide useful insight in this direction. 

Our results are relevant for spin configurations, domain walls and 
magnetic phase transitions in multiferroic materials \cite{fiebig,curnoe}; 
periodic domain walls are yet to be observed in the hexagonal multiferroics 
\cite{fiebig}.  Similarly, our solutions are important for understanding  
structural phase transitions in ferroelectrics \cite{aubry,abel} and 
elastic materials \cite{das}, biophysics problems such as multilamellar 
lipid systems \cite{jcp} as well as field theoretic contexts \cite{raja,pla}. 
These ideas and exact solutions can be generalized to other coupled 
models such as $\phi^6$ (for first order phase transitions) and will 
be discussed elsewhere \cite{phi6}.  
 
\section{Acknowledgment}
A.K. acknowledges the hospitality of the Center for Nonlinear
studies at LANL.  This work was supported in part by the U.S.
Department of Energy.

\end{document}